\documentclass[%
 aip,
 amsmath,amssymb,
 reprint,%
]{revtex4-1}

\usepackage{graphicx}% Include figure files
\usepackage{dcolumn}% Align table columns on decimal point
\usepackage{bm}% bold math
\usepackage[utf8]{inputenc}
\usepackage[T1]{fontenc}
\usepackage{mathptmx}
\usepackage{etoolbox}
\usepackage{url}
\usepackage[final]{changes}

\makeatletter
\def\@email#1#2{%
 \endgroup
 \patchcmd{\titleblock@produce}
  {\frontmatter@RRAPformat}
  {\frontmatter@RRAPformat{\produce@RRAP{*#1\href{mailto:#2}{#2}}}\frontmatter@RRAPformat}
  {}{}
}%
\makeatother
\begin{document}

\preprint{AIP/123-QED}

\title[This article has been accepted by Appl. Phys. Lett. After it is published, it will be found at \url{https://pubs.aip.org/aip/apl}.]{Broadband forward scattering of light by plasmonic balls: role of multipolar interferences}
% Force line breaks with \\
\author{Ranjeet Dwivedi}
\affiliation{ENSEMBLE3 Centre of Excellence, Wolczynska 133, 01-919 Warsaw, Poland}
\author{Maeva Lafitte}
\author{Lionel Buisson}
\author{Olivier Mondain-Monval}
\author{Virginie Ponsinet}
\affiliation{Univ. Bordeaux, CNRS, CRPP, UMR 5031, F-33600 Pessac, France}
 %\altaffiliation[Also at ]{Physics Department, XYZ University.}%Lines break automatically or can be forced with \\
\author{Alexandre Baron}%
 \email{alexandre.baron@u-bordeaux.fr}
\affiliation{Univ. Bordeaux, CNRS, CRPP, UMR 5031, F-33600 Pessac, France}
\affiliation{Institut Universitaire de France, 1 rue Descartes, 75231 Paris Cedex 05, France}%

\date{\today}% It is always \today, today,
             %  but any date may be explicitly specified

\begin{abstract}
Efficient and broadband forward-scattering is a property of prime importance for meta-atoms if they are to be used in self-assembled metasurfaces. Strong contenders include colloidal nanoresonators with tailored multipolar content to achieve the proper interferences that suppress back-scattering. We consider dense plasmonic balls composed of more than a hundred silver nanoinclusions. Numerical simulations provide a full understanding of the role played by multipole moments in the scattering behavior. They are fabricated using emulsion drying and characterized optically. Strong and efficient forward-scattering is demonstrated over the entire visible range. Electric and magnetic dipole resonances of equal amplitude and phase are evidenced. Such plasmonic balls could be used as meta-atoms for bottom-up metasurface applications.
\end{abstract}

\maketitle

\noindent We define plasmonic balls as dense spherical assemblies of plasmonic nanoparticles \cite{rockstuhl2007design}. They exhibit remarkable absorption and scattering properties because of the resonant nature of the inclusions that compose the ball \cite{turek2016self,dwivedi2022electromagnetic}. They belong to the wider class of photonic balls in which we also find systems composed of dielectric nanoparticles \cite{yazhgur2022inkjet}. These dielectric inclusions are usually larger than their plasmonic counterpart and have garnered a lot of interest recently for their ability to produce strong structural coloration \cite{park2014full,vogel2015color}. The case of plasmonic balls is different because the inclusions are very subwavelength and resonant. The absorption and scattering of light by dense plasmonic balls is hard to describe theoretically, because even though the inclusions are subwavelength, the ensemble properties are radically different to those of the constitutive elements and cannot be predicted by traditional effective medium theories because of strong electromagnetic interactions. Recent work has shown empirically that plasmonic balls made of gold nanospheres may act as Huygens meta-atoms \cite{elancheliyan2020tailored}. Huygens meta-atoms are subwavelength structures that act as strong forward-scatterers. They are named in reference to the Huygens-Fresnel principle that states that a propagating plane wave front is equivalent to a continuum of point sources of hemispherical waves emitted in the direction of propagation. These balls can be conveniently assembled through an emulsion route and as a result, they come in large numbers in the form of a suspension or ink that could eventually be used as a coating material for a surface, or even serve as the basic building block of a self-assembled metasurface. 

In this work, high precision T-Matrix numerical simulations are made to design dense silver plasmonic balls in an experimentally achievable situation with the aim of achieving broadband and efficient forward-scattering in the visible range. The simulations reveal the origin of the scattering behaviour and show how they are the result of multipolar interferences between the electric and magnetic dipole moments on the red-end of the visible spectrum and between the electric dipole, magnetic dipole and electric quadrupole moments on the blue-end of the spectrum. Following an emulsion route similar to that proposed by Elancheliyan et al. \cite{elancheliyan2020tailored}, we self-assemble dense plasmonic balls made of silver nanoinclusions and show that they exhibit strong and broadband forward-scattering as designed numerically. This is evidenced by measuring the angular dependence of the light scattered by a suspension of such balls. We also mention differences in the measured and simulated observables.

\begin{figure*}[ht!]
	\centering	
    \includegraphics[width=0.85\textwidth]{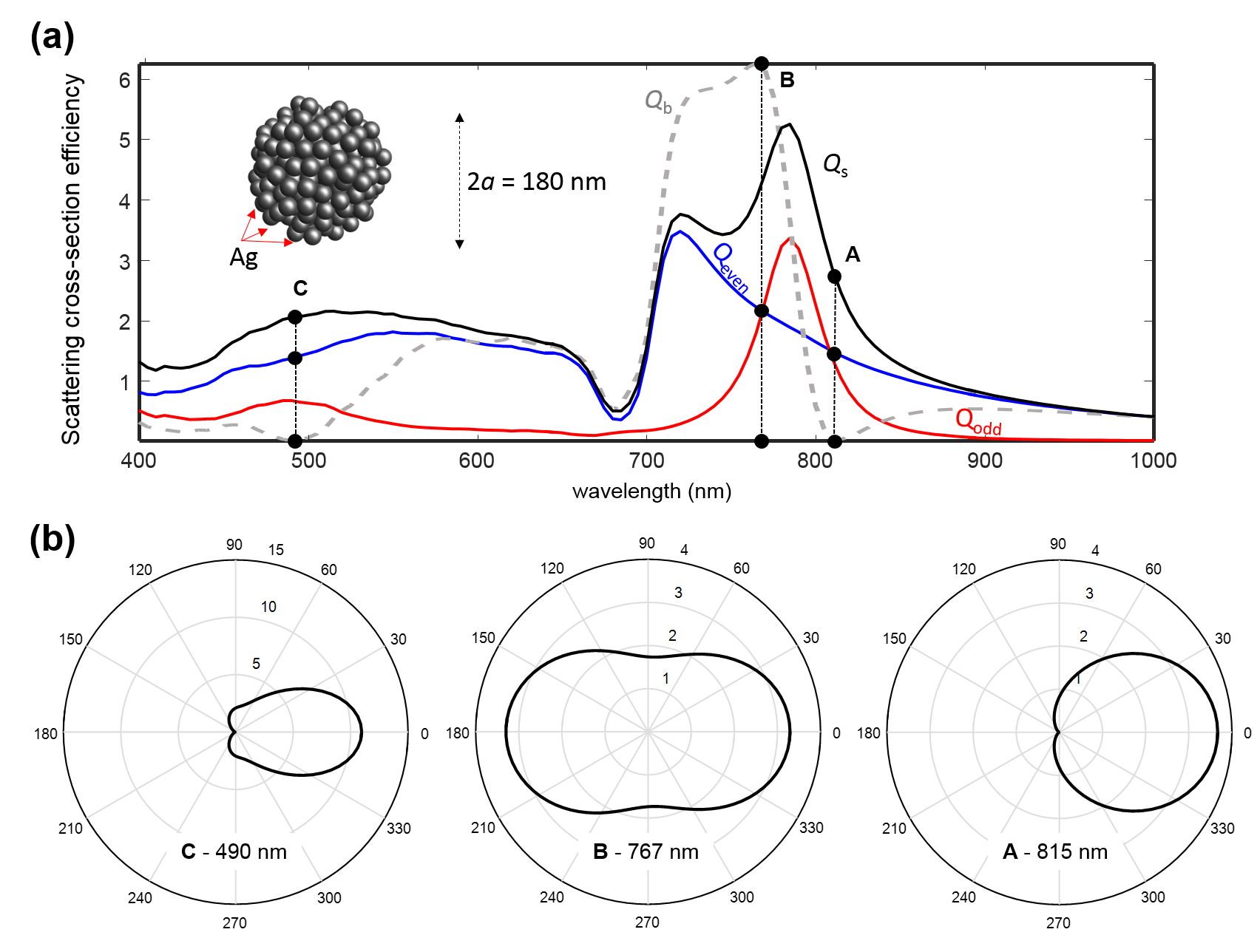}
	\caption{Scattering behavior of silver plasmonic balls. The plasmonic ball has a size $2a=180$ nm and is composed of nanoinclusions of 12 nm in radius. The refractive index of the host medium is 1.42, while the homogeneous matrix surrounding the ball has a refractive index of 1.4. (a) Spectral distribution of the total scattering cross-section efficiency (in black). The blue (red) curve is the superposition of the efficiencies of all even (odd) multipoles. The grey dashed curve is the efficiency for back-scattering. (b) Scattering diagram at three different wavelengths (A: 815 nm, B: 767 nm, C: 490 nm), taken in the plane orthogonal to the polarization direction. The impinging plane wave propagates from left to right.}
 \label{Fig1}
\end{figure*}
\begin{figure*}[t!]
	\centering	
\includegraphics[width=0.95\textwidth]{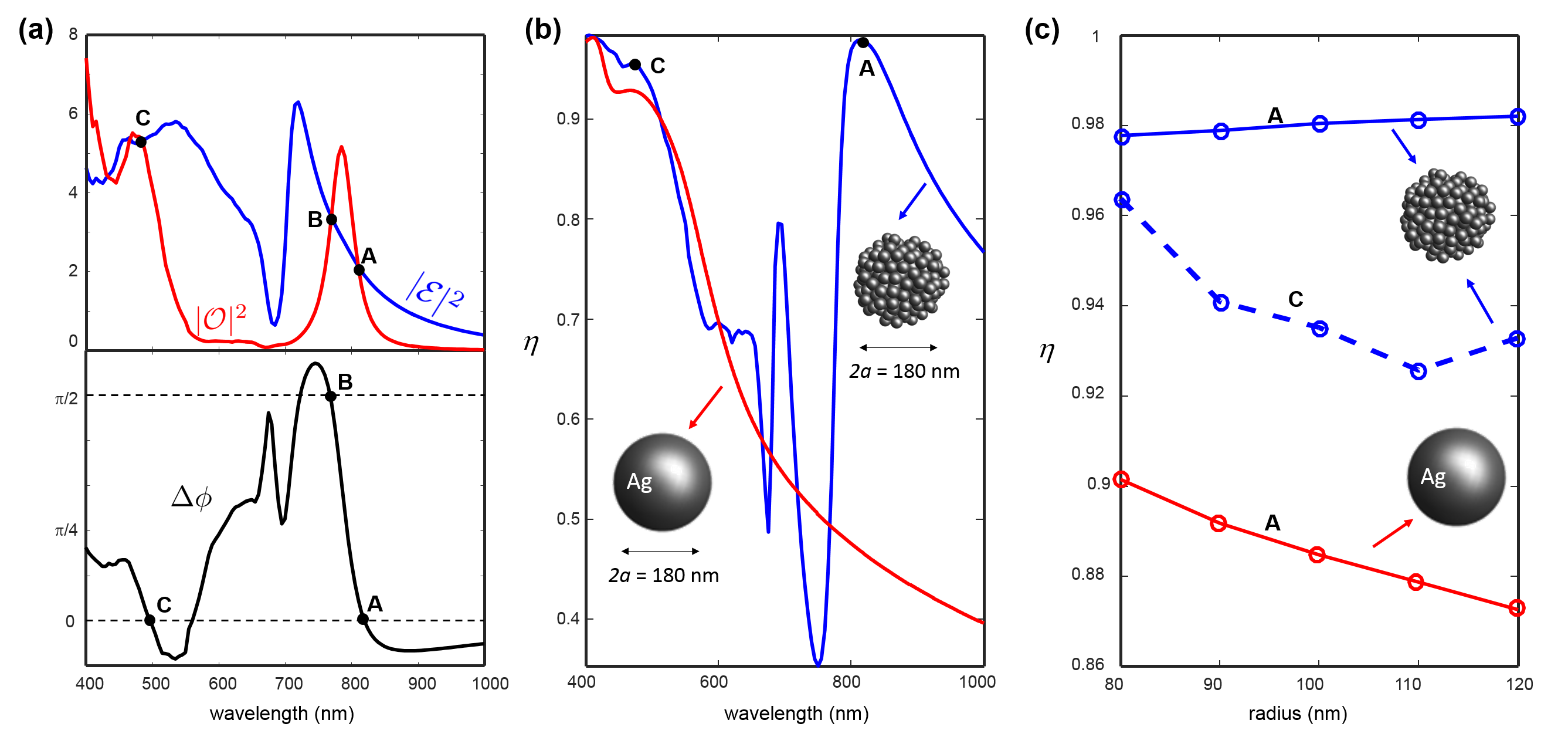}
	\caption{Origin of the scattering behavior and forward-scattering efficiency. (a)  Top panel: Spectral variations of $\lvert\mathcal{E}\rvert^2$ and $\lvert\mathcal{O}\rvert^2$. Bottom panel: Phase difference $\Delta\phi = \arg(\mathcal{O}\mathcal{E}^*)$ between odd and even multipole superpositions. (b) Comparison of the fraction of energy scattered in the forward direction $\eta$ as a function of wavelength for the plasmonic ball of size $2a = 180$ nm (in blue) and a silver nanoparticle of same size (in red). (c) Comparison of $\eta$ for the plasmonic ball (in blue) and the silver nanoparticle (in red) as a function of particle radius at three different wavelengths (A: 815 nm, C: 490 nm).}\label{Fig2}  
\end{figure*}

The plasmonic ball considered for the numerical simulations consists of a sphere of radius $a$ in which silver nanoparticles of radius $r=12$ nm are packed (see illustration in inset of Fig. \ref{Fig1}(a)). The host medium surrounding the ball has a refractive index $n_h = 1.42$. Fixing $a$ and the volume fraction $f$ sets the number of inclusions $N$ per ball. \added{In the end, the radius $a$ of the ball is the central distance between two farthest inclusions within the ball and the diameter of an inclusion. In other words it is the maximal Feret radius}. We chose $f=0.4$. The dielectric constant of silver is taken from Jonhson and Christy \cite{Johnson1972}. For the computation, the ball is numerically generated by randomly distributing the inclusions within a cubic box of length $L$ larger than $a$ using the Lubachevsky-Stillinger (LS) algorithm \cite{lubachevsky1990geometric}. \added{The LS-algorithm is a means of generating a random distribution of spheres in a box with a targeted volume fraction consistent with the experimental non-crystalline distributions.} The number of inclusions of radius $r$ is $N = 168$.

The resulting ball is then numerically implemented to compute its scattering behavior using the multiple sphere T-matrix software developed by Mackowski \cite{winnt}, which enables us to compute the multipole Mie coefficients for wavelengths between 400 and 1000 nm. \added{For $f=0.4$, full-wave simulations such as the T-matrix are necessary to accurately predict the behaviour of the ball because classical effective medium theories such as the extended Maxwell-Garnett theory are highly inaccurate for such dense plasmonic media \cite{dwivedi2022electromagnetic}. We note however that forward scattering can be described by an effective homogeneous sphere in the Maxwell-Garnett approximation for photonic balls made of low index dielectric inclusions \cite{yazhgur2021light}.} The multipole Mie coefficients are the electric and magnetic mutipole moments ($a_n$ and $b_n$ respectively). The process of generating a ball is repeated $P$ times and each multipole coefficient is averaged over the $P$ realizations until $\langle a_n\rangle$ and $\langle b_n\rangle$ are converged within an accuracy well below one percent. \added{Once averaged over sufficiently many realizations, we hypothesize that the scattering properties are not contingent on the method employed to generate the plasmonic ball}. We find that $P=50$ is sufficient to reach such an accuracy. Multipoles up to order $n=3$ are kept, all higher-order multipoles are found to be vanishingly small. Next the scattering cross-section efficiency $Q_s = Q_\mathrm{even}+Q_\mathrm{odd}$ is computed and decomposed into even and odd contributions
\begin{eqnarray}
    Q_\mathrm{even} &=& \frac{2}{x^2}\sum_{n}\left[(4n-1)\lvert a_{2n-1}\rvert^2+(4n+1)\lvert b_{2n}\rvert^2\right]\\
    Q_\mathrm{odd} &=& \frac{2}{x^2}\sum_{n}\left[(4n-1)\lvert b_{2n-1}\rvert^2+(4n+1)\lvert a_{2n}\rvert^2\right]
\end{eqnarray}
where $x = 2\pi n_h a /\lambda$ is the reduced frequency.

Figure \ref{Fig1}(a) displays $Q_s(\lambda)$. The contributions to the scattering efficiency from even and odd multipoles is also shown in blue and red respectively. We find three points of interest that correspond to remarkable points of the spectral variations of the efficiency for back-scattering (dashed gray curve). Point A is a minimum of back-scattering, which occurs at $\lambda = 815$ nm, when the scattering of even and odd multipoles is equal both in amplitude and in phase. This fact is evidenced by the crossing of the blue and red curves passed the resonance of both superpositions. The corresponding radiation diagram (see right-hand polar plot in Fig. \ref{Fig1}(b)), taken in the plane parallel to the incident wavevector and orthogonal to the polarization direction, exhibits a clear zero at a scattering angle $\theta = 180^\circ$ and is essentially directed in the forward direction. The impinging wavevector is directed from left to right. Point B is a maximum of back-scattering at $\lambda = 767$ nm. This time the radiation diagram displays a radiation pattern that is balanced between the backward and forward directions (see central polar plot in Fig. \ref{Fig1}(b)). In this situation the even and odd scattering have equal amplitudes but are dephased by $\pi/2$ as will be evidenced further down. Finally, point C is another minimum of back-scattering at $\lambda = 490$ nm, with a radiation diagram that also presents a zero of scattering at $\theta = 180^\circ$, albeit with a distribution of the scattering that is more directional. Here, the scattering from even and odd multipoles are not equal in amplitude. This is due to the fact that higher order multipolar interferences occur, as we shall now see.

\begin{figure}[b!]
	\centering	
\includegraphics[width=0.95\columnwidth]{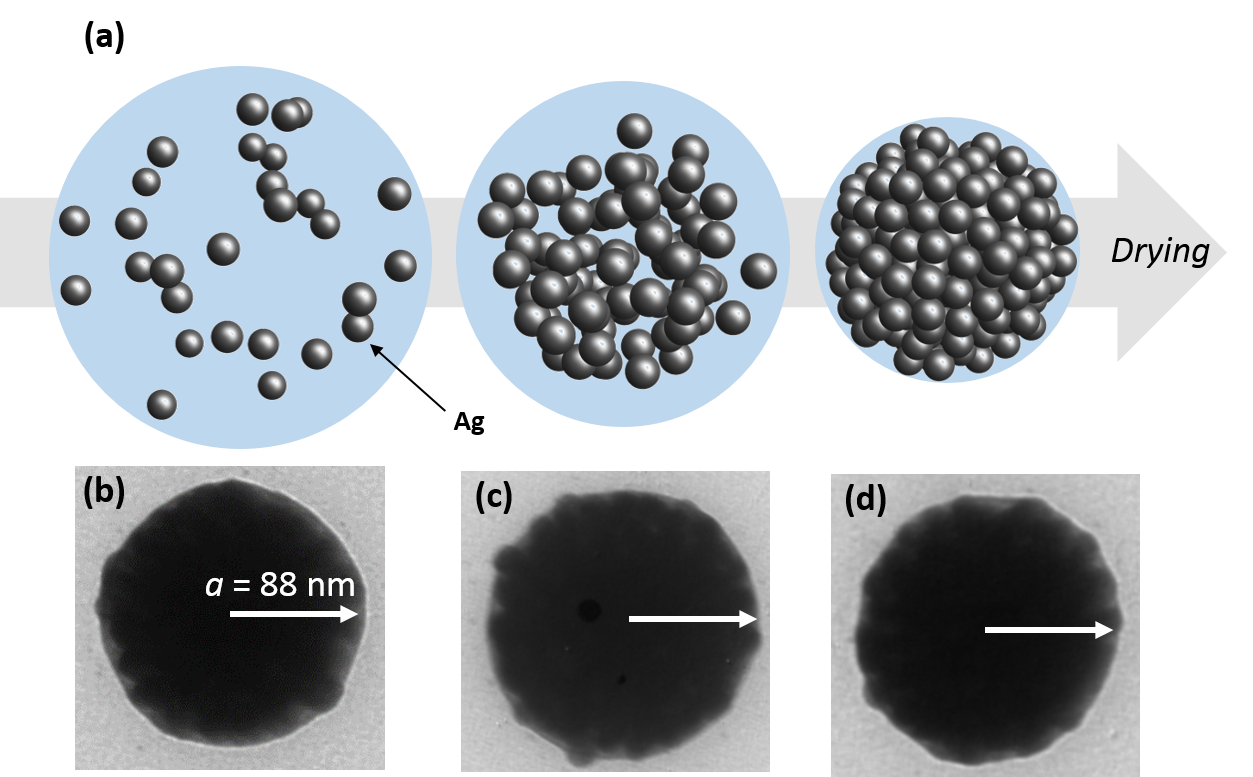}
	\caption{Experimental plasmonic balls. (a) Illustration of the emulsion drying process of plasmonic ball formation from an emulsion of dilute silver nanoparticle suspensions to a dense spherical ball. (b) Transmission electron micrographs of typical plasmonic balls obtained experimentally. The arrow is $88$ nm in length corresponding to the average hydrodynamic radius of the balls in solution.}\label{Fig3}  
\end{figure}
\begin{figure*}[t!]
	\centering	
\includegraphics[width=0.95\textwidth]{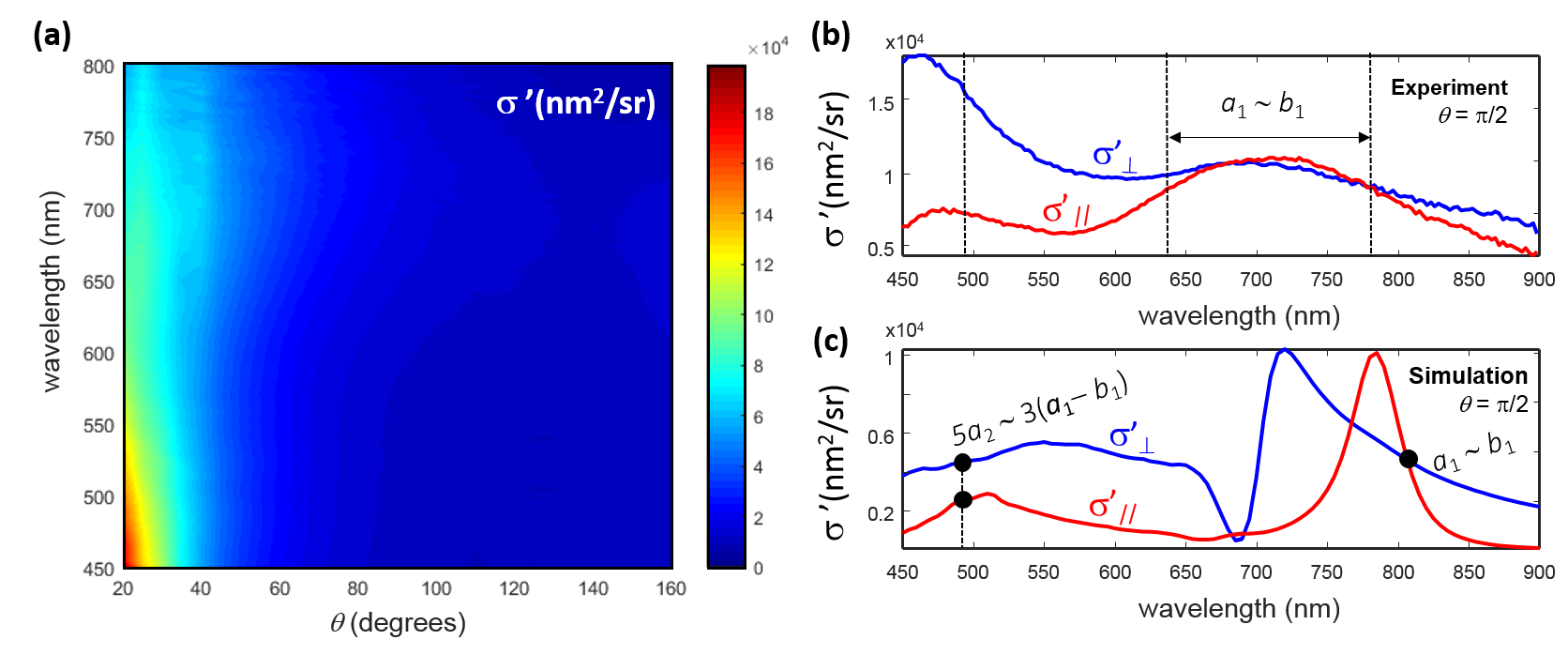}
	\caption{Forward-scattering of silver plasmonic balls. (a) Color plot of the differential scattering cross-section $\sigma'(\theta,\lambda)$ (in nm$^2$/sr) as a function of wavelength and scattering angle. (b) Spectral variations of the parallel (in blue) and orthogonal (in red) components of $\sigma'(\theta=\pi/2,\lambda)$. The curves are normalized to the maximum value of the parallel component of $\sigma'$. (c) Simulated spectral variations of the perpendicular (in blue) and parallel (in red) components of $\sigma'$.}\label{Fig4}
\end{figure*}
To get a clearer understanding of the cancellation of the back-scattering, it is useful to consider the efficiency for back-scattering $Q_b = \lvert\mathcal{E}-\mathcal{O}\rvert^2/x^2$, expressed as the difference between even and odd multipolar superpositions \cite{bohren2008absorption}
\begin{eqnarray}
    \mathcal{E} &=& \sum_{n} \left[(4n-1)a_{2n-1}+(4n+1) b_{2n}\right]\\
    \mathcal{O} &=& \sum_{n} \left[(4n-1)b_{2n-1}+(4n+1) a_{2n}\right]
\end{eqnarray}
The spectral variations of $\lvert \mathcal{E}\rvert^2$ and $\lvert \mathcal{O}\rvert^2$ are shown on the top panel of Fig. \ref{Fig2}(a), while the phase difference between the two $\Delta\phi = \arg(\mathcal{O}\mathcal{E}^*)$ is shown on the bottom panel. We see that the remarkable points marked in Fig. \ref{Fig1} coincide with specific interferences between odd and even multipole superpositions. Indeed, at point A, they are equal in amplitude and in phase ($\Delta\phi = 0$) resulting in a zero of $Q_b$. As a matter of fact at this point, the plasmonic ball is almost completely dipolar ($n>1$ terms are negligible) and as a result, the first Kerker condition is satisfied: $a_1 = b_1$ \cite{kerker1983electromagnetic}. The electric ($a_1$) and magnetic ($b_1$) dipoles constructively (resp. destructively) interfere in the forward (resp. backward) direction. For point B, even and odd contributions have equal amplitudes but are dephased ($\Delta\phi = \pi/2$) and we have: $\mathcal{E}=\mathcal{O}^*$. Since at this point too, the system is essentially dipolar, this translates into $a_1 = b_1^*$. So this time the backward radiation is balanced with that in the forward direction (see online supplementary material). Point C has more multipolar content as the electric quadrupole ($a_2$) can no longer be neglected. We have $\vert\mathcal{E}\rvert = \lvert\mathcal{O}\rvert$ and $\Delta\phi = 0$, which translates into an interference between the electric dipole on the one-hand and the magnetic dipole and electric quadrupole on the other: $3a_1 = 3b_1+5a_2$. \added{The phase differences observed here are a consequence of the plasmonic interactions within the plasmonic ball. They stem from an interplay between the volume fraction of inclusions within the ball, its radius and the inclusion radius. A comprehensive understanding of this interplay remains outside the scope of this study. Here we focus only on revealing the origin of the forward scattering behaviour we wish to target experimentally.}
The efficiency of the forward scattering can be described by the figure-of-merit $\eta$ which is the fraction of energy scattered in the forward direction ($\lvert\theta\rvert<90^\circ$). $\eta(\lambda)$ for the plasmonic ball is compared to that of a silver nanoparticle of same size on Fig. \ref{Fig2}(b). We see that the plasmonic ball does significantly better in terms of forward-scattering for $\lambda > 770$ nm, reaching a maximum of 98\% of the total scattered energy radiated in the forward direction in A. For $\lambda<600$ nm, we see that both particles display a similar figure-of-merit. In C, where the electric quadrupole is present in the multipolar content, $\eta \approx 96\%$.

The observations made on the plasmonic ball of radius $a=90$ nm were also observed for simulations of balls of radius $a\in[80,90,100,110,120]$ nm and identical volume fraction ($f=0.4$). Figure \ref{Fig2}(c) compares $\eta(a)$ at point A, defined as the first minimum of $Q_b$ (starting from the long-wavelength limit), for both the plasmonic ball and the silver nanoparticle. We see that at this point, $\eta>97\%$ for all sizes of the plasmonic ball and it scatters 9 to 10$\%$ more energy forward compared to the silver nanoparticle of same size. 

The plasmonic balls are fabricated via an emulsion route sketched on Fig. \ref{Fig3}(a). An initial aqueous suspension of silver nanoparticles coated with PVP of radius $r = 12$ nm were emulsified in an oil phase composed of dodecane oil of refractive index 1.42. The internal water phase is then evaporated using a rotavapor flask at room temperature under low pressure. The samples are characterized using dynamic light scattering and the mean hydrodynamic radius is found to be $a = 88$ nm with a relative standard deviation of 8$\%$. Typical transmission electron micrographs (TEM) of the plasmonic balls found in the emulsion are shown on Fig. \ref{Fig3}(b-d). They are typically composed $\sim$ 150 silver nanoinclusions. \added{Small-angle X-ray scattering experiments performed on the fabricated sample indicate no evidence of crystallization of the NPs within the balls, we thus conclude that the plasmonic balls are disordered (see supplementary material -- SM -- for additional information on fabrication and structural characterization of the plasmonic balls).}

We characterize the suspension obtained, by using a variable-angle spectroscopic polarization resolved static light scattering setup (described in section 3.2 in \cite{aradian}). The sample is mounted on a goniometric mount so as to measure the light scattered by the sample over a wide range of scattering angles from $\theta = 20$° (forward) to $\theta = 160$° (backward). The light scattered is analyzed on a detection arm that rotates around the sample. It contains an analyzer which selects either the polarization parallel or orthogonal to the scattering plane. As a result, the differential scattering cross-section $\sigma'(\theta,\lambda)$ can be retrieved, as the sum of the contributions from polarizations parallel ($\sigma'_{\parallel}$) and orthogonal ($\sigma'_\perp$) to the scattering plane, that can both be quantitatively determined from proper normalization and knowledge of the concentration of plasmonic balls in solution. $\sigma'(\theta,\lambda)$ is displayed as a color plot on Fig. \ref{Fig4}(a) and shows that the scattering from a suspension of plasmonic balls is dually resonant at $\lambda\approx 700$ nm and $\lambda\approx450$ nm, reminiscent of points A, where $a_1 \approx b_1$ and C, where $5a_2 \approx 3(a_1-b_1)$. Furthermore the scattering is essentially directed in the forward direction within a scattering angle smaller than 45$^\circ$ typically. The online SM provides details on how $\sigma'$ could be determined quantitatively. Both parallel and perpendicular contributions to $\sigma'$ are shown as a function of wavelength for $\theta = \pi/2$ on Fig. \ref{Fig4}(b) and compared to the simulation on Fig. \ref{Fig4}(c). At that scattering angle, the perpendicular and parallel contributions only contain a superposition of even and odd multipoles respectively \cite{bohren2008absorption}. We see that $\sigma'_{\parallel}\approx\sigma'_{\perp}$ experimentally for $\lambda\in[650,780]$ nm in a spectral region where it is mostly dipolar. We can reasonably infer that for such wavelengths, $a_1\approx b_1$. There are notable differences with the simulated spectrum, where $a_1 = b_1$ only at point A. The experimental spectrum is smeared in comparison to theory and the overlap occurs over a wider spectral range. We attribute this difference to the polydispersity of the suspension of plasmonic balls. Nonetheless it is remarkable that the intervals of variation of $\sigma'$ both in the experiment and the theory are very close, notably near the resonance ($\approx 1\times10^4$ nm$^2$/sr), keeping in mind that this result is retrieved with no adjustable parameters. For $\lambda=490$ nm, corresponding to point C in the simulation, where the electric quadrupole interferes with the electric and magnetic dipoles, we note that the fraction $\sigma'_\perp/\sigma'_\parallel$ are sensibly the same and close to 0.5 (0.45 in the experiment and 0.56 in the simulation). 

We have simulated and fabricated dense plasmonic balls and demonstrated that they exhibit broadband and efficient resonant forward-scattering at visible frequencies. We reveal that this behavior is the result of multipolar interferences and highlight the role of the electric quadrupole in the blue-end of the spectrum and that of the magnetic dipole at the red-end. These plasmonic balls come in large quantities in solution and can be scaled-up easily. \added{However, as evidenced in this letter, polydispersity should be reduced for future functional requirements and device reliability. To this end, on chip microfluidic approaches appear as promissing solutions as they are known to produce emulsions with very well controlled droplet sizes. They hold the potential of drastically reducing polydispersity. Furthermore, an in-depth study  of the impact of plasmonic ball size and volume fraction  could also reveal how polydispersity influences its multipolar content and spectral features. When such issues are solved, plasmonic balls will have the potential} to serve as efficient meta-atoms for metasurfaces applications, notably those that require strongly forward-scattering resonators also known as generalized Huygens sources \cite{staude2013tailoring,kruk2016invited,dezert2019complete}.  
\vspace{-15pt}

\section*{SUPPLEMENTARY MATERIAL}
%\vspace{-15pt}
\noindent \added{A supplementary material file is available online. It contains additional information on the general theoretical framework used in this work, numerical implementation of the plasmonic ball, fabrication and structural characterization of the plasmonic balls, definition of the fraction of energy scattered forward, as well as a description of the experimental setup and method used for the optical characterization of the balls and retrieval of the differential scattering cross-section.} 

\section*{AUTHOR DECLARATIONS}
%\vspace{-15pt}
\noindent The authors have no conflicts to disclose. 

\section*{DATA AVAILABILITY STATEMENT}
%\vspace{-10pt}
\noindent The data that support the findings of this study are available from the corresponding author upon reasonable request.

%\nocite{*}

\section*{REFERENCES}
\vspace{-10pt}
\bibliography{aipsamp}% Produces the bibliography via BibTeX.

\end{document}